\documentclass[aps,pre,showpacs,groupedaddress]{revtex4}

\usepackage[dvips]{graphicx}
\usepackage{amssymb}
\usepackage{latexsym}
\usepackage{pifont}
\usepackage{wasysym}

\begin{document}

\title{Influence of humidity on granular packings with moving walls}

\author{Y. Bertho\footnote{Present address: Center for Nonlinear
Phenomena and Complex Systems, and Microgravity Research Center,
Universit\'e Libre de Bruxelles, Campus Solbosch, CP 165/62, 1050
Brussels, Belgium}, Th. Brunet\footnote{Present address:
Laboratoire LMPDI, UMR 8108, Cit\'e Descartes, B\^atiment
Lavoisier, Universit\'e de Marne la Vall\'ee, 77454 Marne la
Vall\'ee Cedex 2, France}, F. Giorgiutti-Dauphin\'e and J.-P.
Hulin}

\affiliation{Laboratoire FAST, UMR CNRS 7608, B\^at.~502,
Universit\'e Paris XI, 91\,405 Orsay Cedex, France}


\begin{abstract}
A significant dependence on the relative humidity $H$ for the
apparent mass ($M_\mathrm{app}$) measured at the bottom of a
granular packing inside a vertical tube in relative motion is
demonstrated experimentally. While the predictions of Janssen's
model are verified for all values of $H$ investigated ($25\%\leq
H\leq 80\%$), $M_\mathrm{app}$ increases with time towards a
limiting value at high relative humidities ($H\ge 60\%$) but
remains constant at lower ones ($H=25\%$). The corresponding
Janssen length $\lambda$ is nearly independent of the tube
velocity for $H\ge 60\%$ but decreases markedly for $H=25\%$.
Other differences are observed on the motion of individual beads
in the packing: for $H=25\%$, they are almost motionless while the
mean particle fraction of the packing remains constant; for $H\ge
60\%$ the bead motion is much more significant and the mean
particle fraction decreases. The dependence of these results on
the bead diameter and their interpretation in terms of the
influence of capillary forces are discussed.
\end{abstract}

\maketitle

Dense granular flows in vertical channels are encountered in many
industrial processes and often display intermittency or blockage
effects representing important practical problems \cite{Laouar98,
Bertho02, Bertho03b}. These effects depend on the force
distribution in the moving grain packing and on its interaction
with the walls: these depend in turn on the relative
humidity $H$ of the atmosphere. In static packings,
capillary forces strongly influence, for instance, the stability
of sand piles as a function of $H$ \cite{Albert97, Bocquet98,
Fraysse99, Hornbaker97, Mason99, Tegzes99}. In the present work,
the influence of humidity on force transmission in a granular
packing inside a vertical tube in relative motion is analyzed from
variations of its apparent weight measured at the bottom of the
packing.

Stress transmission in static or quasistatic granular packings has
been frequently investigated theoretically and experimentally
\cite{Bouchaud95, Mounfield96, Mueth98, Pitman98, Vanel00,
Ovarlez03, Ovarlez03b, Bertho03, Landry03, Landry03b} following
the pioneering work and Janssen \cite{Janssen95}. This latter
study and subsequent ones predict that the vertical stress at the
bottom of a grain packing inside a vertical cylindrical tube
reaches exponentially a limiting value as the height of the
packing increases. This is due to the redirection of vertical
stresses towards the side walls where friction between grains and
the walls occurs: this shields the lower grain layers from the
weight of the upper ones so that the apparent mass
$M_\mathrm{app}$ measured at the bottom is lower than the total
mass $M$ of the grains.

In a recent paper \cite{Bertho03} we demonstrated that, for a
constant relative humidity $H=50\%$, Janssen's predictions remain
valid for a grain packing inside a vertical tube moving upwards
relative to the grains at velocities between $10^{-2}$ and
$35\,mm\,s^{-1}$. The present study uses these results as a tool
to analyse the dependence of stress transmission on the relative
humidity $H$ in the range $25\%\leq H \leq 80\%$: $H$ will be
shown to influence strongly the variations and value of the
apparent mass of the grains during the tube motion. Analyzing the
motion of the grains during the tube displacement and the
influence of the bead diameter confirms this influence.

\begin{figure}[ht!]
\includegraphics[width=8cm]{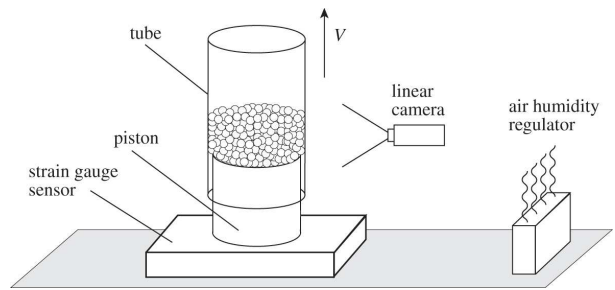} \caption{Schematic view of
the experimental set-up.} \label{setup}
\end{figure}

The experimental setup includes a vertical glass tube (length =
$400\,mm$, internal diameter $D=30\,mm$) containing a mass $M$ of
monodisperse glass beads (diameter $d$, density $\rho=2.53\times
10^3\,kg\,m^{-3}$). The tube can be moved at velocities $V$
between $10^{-1}$ to $130\,mm\,s^{-1}$ over distances of $110\,mm$
with accelerations up to $0.7\,m\,s^{-2}$. The acceleration and
deceleration distances are of order $12\,mm$ at the fastest
velocity and of less than $10^{-1}\,mm$ for $V\le 10\,mm\,s^{-1}$.
A static cylindrical piston (diameter $29\,mm$) is inserted at the
bottom of the tube without any mechanical contact
(Fig.~\ref{setup}); it is screwed onto a sensor measuring the
force applied by the packing on the piston. The apparent mass of
the grains is determined up to 256 times per second with an
uncertainty of $\pm 0.1\,g$. The relative humidity $H$ is
regulated to within $\pm 2\%$.

\begin{figure}[b!]
\includegraphics[width=8cm]{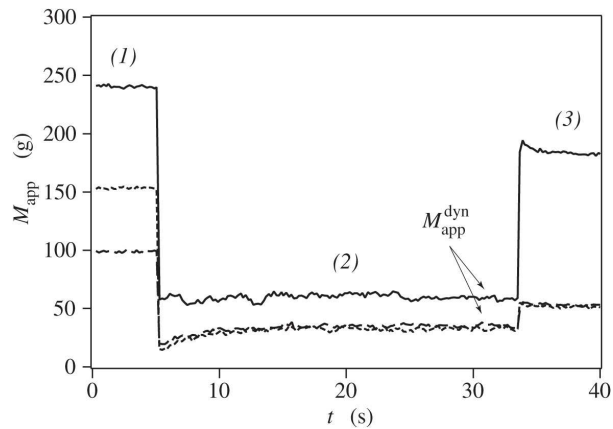} \caption{Variation of the
apparent mass $M_\mathrm{app}$ of a grain packing ($d=2\,mm$) of
total mass $M=300\,g$ as a function of time: ({\it 1})~right after
filling the tube, ({\it 2})~tube moving upwards
($V=4\,mm\,s^{-1}$), ({\it 3})~tube stopped. The curves correspond
to three different relative humidities: (---)~$H=25\%$,
(--~--)~$H=60\%$, ({\scriptsize -~-~-})~$H=80\%$.} \label{comparH}
\end{figure}

Figure~\ref{comparH} displays variations of the apparent mass
$M_\mathrm{app}$ as a function of time for three different
relative humidities $H=25\%$, $60\%$ and $80\%$ and a same
diameter $d=2\,mm$ of the beads. The total mass of grains is
$M=300\,g$ in all experiments (the packing height is $\simeq
270\,mm$). As expected from Janssen's model, the apparent mass
after filling the tube [domain {\it (1)} in Fig.~\ref{comparH}] is
lower than $M$. Its value $M_\mathrm{init}$ varies strongly from
one experiment to another (typ. $\pm 50\%$ of the mean value) due
to the variability of the force chains redirecting the weight
towards the side walls.
As the tube starts to move upwards, $M_\mathrm{app}$ decreases
suddenly. In contrast with the value of $M_\mathrm{init}$, the
variation of $M_\mathrm{app}$ during the motion [domain {\it (2)}
in Fig.~\ref{comparH}] is very reproducible from one experiment to
another~\cite{Bertho03} at all relative humidities $H$ (typical
deviation $\pm 5\%$). These variations of $M_\mathrm{app}$ with
time depend however on $H$ (Fig.~\ref{comparH}).

For $H=60\%$, $M_\mathrm{app}$ increases towards a limiting value
$M_\mathrm{app}^\mathrm{dyn}$ while the tube moves.
$M_\mathrm{app}^\mathrm{dyn}$ is reached over a distance
independent of the tube velocity $V$. This result coincides with
that reported previously for $H=50\%$ \cite{Bertho03}. The curves
are similar for $H=80\%$ (Fig.~\ref{comparH}) and, at a given $V$,
variations for $H=60\%$ and $80\%$ overlay well (note that the
difference between the values of $M_\mathrm{init}$ for $H=60\%$
and $80\%$ are not meaningful and correspond to the fluctuations
pointed above). For $H=25\%$, $M_\mathrm{app}$ still drops
abruptly after the tube starts to move but remains practically
constant thereafter. This constant value is also larger than that
of $M_\mathrm{app}^\mathrm{dyn}$ for $H=60\%$ and $80\%$.

After the tube has stopped [domain {\it (3)} in
Fig.~\ref{comparH}], $M_\mathrm{app}$ increases again for all
relative humidities $H$: the final apparent mass
$M_\mathrm{final}$ remains however lower than the initial one
$M_\mathrm{init}$ and the dispersion of its values is
significantly larger than in domain {\it (2)}.

These results indicate that lowering $H$ influences the dynamics
of the rearrangement of the force chains in the packing when the
walls are set in motion. In order to study quantitatively these
effects, we checked that Janssen's predictions were verified for
all $H$ values. This was realized by studying variations of the
limiting value $M_\mathrm{app}^\mathrm{dyn}$ as a function of the
total mass $M$. The model assumes that the redirection of the
vertical stress $\sigma_{zz}$ into an horizontal component
$\sigma_{rr}$ is characterized by a coefficient $K$ and that the
ratio between the tangential and normal stresses at the walls is
the Coulomb friction coefficient $\mu$. Assuming a zero vertical
stress $\sigma_{zz}$ at the surface of the packing ($z=0$), its
value $\sigma_{zz} (z_0)$ at the bottom ($z=z_0$) verifies:
\begin{equation}
\sigma_{zz}(z_0) = \rho cg\lambda \left (1-e^{-z_0/\lambda}\right
), \label{eq:janssen}
\end{equation}
in which the characteristic length $\lambda$ is given by:
\begin{equation}
\lambda=D / 4\mu K. \label{eq:lambda}
\end{equation}
The effective weight $M_\mathrm{app}g$ measured by the sensor is
equal to $\sigma_{zz}(z_0)\pi D^2/4$ and the total mass $M$ of the
grains is: $\rho c z_0\pi D^2 /4$ ($c$ is the solid volume
fraction of the packing). By combining Eqs.~(\ref{eq:janssen}) and
(\ref{eq:lambda}), one obtains the following expression of the
apparent mass $M_\mathrm{app}$:
\begin{equation}
\frac{M_\mathrm{app}}{M_{\infty}}=1-\exp \left
(-\frac{M}{M_{\infty}}\right ), \label{eq:janssenM}
\end{equation}
where $M_\infty$ is the predicted asymptotic value of
$M_\mathrm{app}$ when $M$ increases:
\begin{equation}
M_\infty = \rho c\lambda \pi D^2 /4.
\label{eq:Minfini}
\end{equation}

\begin{figure}[t!]
\includegraphics[width=\textwidth]{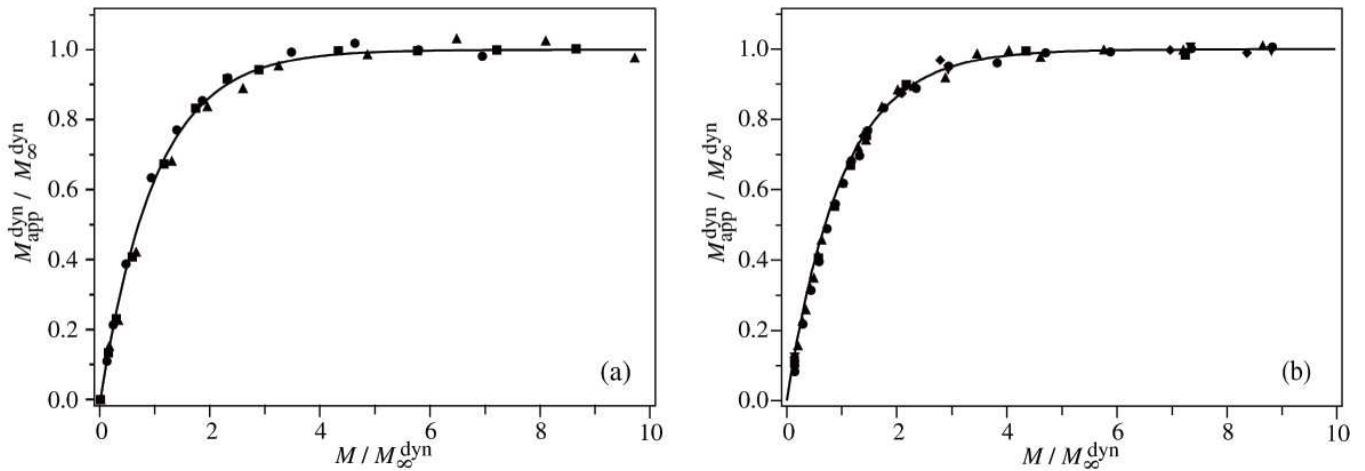} \caption{Reduced
apparent mass $M_\mathrm{app}^\mathrm{dyn}/M_{\infty}$ \emph{vs.}
reduced total mass $M/M_{\infty}$ (bead diameter $d=2\,mm$). The
solid line corresponds to Eq.~(\ref{eq:janssenM}) --- (a) Tube
velocity: $V=4\,mm\,s^{-1}$, relative humidity:
($\bullet$)~$H=25\%$, ($\blacksquare$)~$H=60\%$,
($\blacktriangle$)~$H=80\%$.
--- (b) Relative humidity: $H=60\%$, tube velocity:
($\blacklozenge$)~$V=0.02\,mm\,s^{-1}$,
($\blacksquare$)~$V=0.2\,mm\,s^{-1}$,
($\bullet$)~$V=2\,mm\,s^{-1}$,
($\blacktriangle$)~$V=20\,mm\,s^{-1}$,
($\blacktriangledown$)~$V=34\,mm\,s^{-1}$.} \label{reducedmass}
\end{figure}
In order to verify the validity of relation~(\ref{eq:janssenM}),
the ratio $M_\mathrm{app}^\mathrm{dyn}/M_\infty^\mathrm{dyn}$ is
plotted in Figs.~\ref{reducedmass}a--b as a function of the
reduced total mass $M/M_\infty^\mathrm{dyn}$
($M_\infty^\mathrm{dyn}$ is taken equal to the limiting value of
$M_\mathrm{app}^\mathrm{dyn}$ for large values of $M$). Both for
different relative humidities $H$ (Fig.~\ref{reducedmass}a) and
different tube velocities $V$ (Fig.~\ref{reducedmass}b), the
variations of $H$ coincide precisely with theoretical predictions
from Eq.~(\ref{eq:janssenM}).

Following these results, Janssen's hypothesis can be assumed to be
valid in all our experiments. For given velocity, relative
humidity and bead diameter values, a single experiment using a
large total mass $M$ is required to determine the length
$\lambda$: $\lambda$ can indeed be computed from
Eq.~(\ref{eq:Minfini}) using the asymptotic apparent mass
$M_\infty^\mathrm{dyn}$ measured in this way. The relative error
on the value of $\lambda$ is of the order of $\pm 4\%$.
\begin{figure}[t!]
\includegraphics[width=\textwidth]{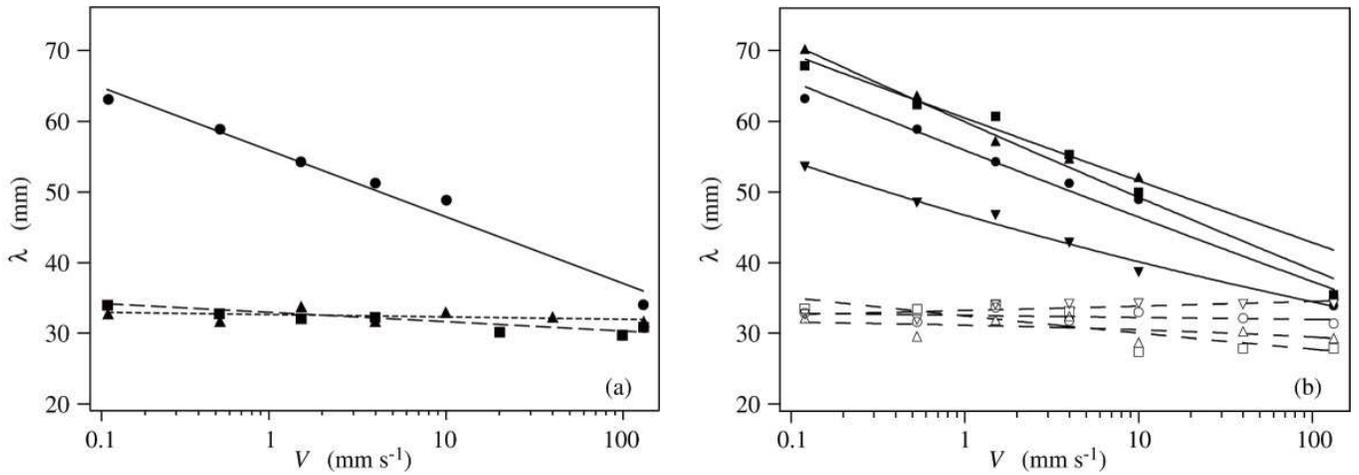} \caption{Variations
of Janssen's length $\lambda$ with the tube velocity $V$ ---
(a)~for relative humidities: ($\bullet$)~$H=25\%$,
($\blacksquare$)~$H=60\%$, ($\blacktriangle$)~$H=80\%$ and bead
diameter $d=2\,mm$. --- (b)~for relative humidities $H=25\%$
(black symbols) and $H=80\%$ (open symbols) and for several beads
diameters: ($\triangledown$,$\blacktriangledown$)~$d=1.5\,mm$,
($\circ$,$\bullet$)~$d=2\,mm$,
($\square$,$\blacksquare$)~$d=3\,mm$,
($\triangle$,$\blacktriangle$)~$d=4\,mm$.} \label{lambdaV}
\end{figure}

The influence of the relative humidity is then characterized by
comparing the variations of $\lambda$ with the tube velocity for
three different values of $H$ (Fig.~\ref{lambdaV}a). For $H=60\%$
and $80\%$, $\lambda$ is nearly constant and close to the tube
diameter $D$ with $\lambda=(32\pm 1.5)\,mm=1.1D$ over a range of
velocities of three decades. On the contrary, for $H=25\%$,
$\lambda$ decreases roughly linearly with $V$ from 65 to $35\,mm$
over the same range of velocities (in this curve, data points for
$V\simeq 50\,mm\,s^{-1}$ could not be obtained due to a mechanical
resonance of the set-up).

Another important parameter is the diameter $d$ of the beads:
variations of $\lambda$ as a function of $V$ are displayed in
Fig.~\ref{lambdaV}b for two different relative humidities
($H=25\%$ and $H=80\%$) and four different diameters $d$ from 1.5
to $4\,mm$.
At a given relative humidity $H$, similar variations of $\lambda$
with the tube velocity $V$ are observed for all bead diameters.
For $H=80\%$, $\lambda$ is constant or varies slowly with $V$,
while it decreases by a factor of 2 for $H=25\%$. However, at
$H=25\%$, the variation is slower for the smallest beads
($d=1.5\,mm$) than for the others.
\begin{figure}[t!]
\includegraphics[width=8cm]{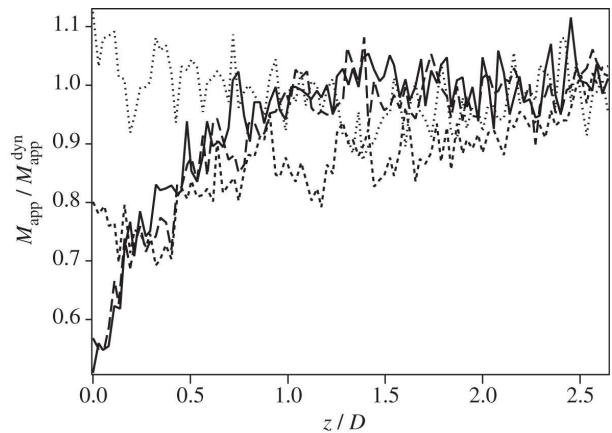} \caption{Variation of
$M_\mathrm{app}/M_\mathrm{app}^\mathrm{dyn}$ with $z/D$ for a tube
velocity $V=4\,mm\,s^{-1}$, a relative humidity $H=60\%$ and bead
diameters: ({\bf{-----}})~$d=1.5\,mm$,
({\bf{---}~{---}})~$d=2\,mm$, $(-~-)$~$d=3\,mm$,
($\cdot$~$\cdot$~$\cdot$)~$d=4\,mm$ ($M=300\,g$).}
\label{fig:mappvsz}
\end{figure}

Variations of the normalized apparent mass
$M_\mathrm{app}/M_\mathrm{app}^\mathrm{dyn}$ with the normalized
displacement $z/D$ provide additional information. Overall, the
same qualitative differences between $H=80\%$ and $25\%$ are
observed for all bead diameters as in Fig.~\ref{comparH} for
$d=2\,mm$. For $H=80\%$, after a sharp initial drop,
$M_\mathrm{app}$ increases slowly towards a limiting value at
nearly all velocities, while it remains roughly constant for
$H=25\%$. For the intermediate value $H=60\%$, the variation may
depend however on the bead diameter. In Fig.~\ref{fig:mappvsz},
for $d=1.5\,mm$ and $d=2\,mm$,
$M_\mathrm{app}/M_\mathrm{app}^\mathrm{dyn}$ increases with $z/D$
towards 1. For the largest beads ($d=4\,mm$), on the contrary,
$M_\mathrm{app}/M_\mathrm{app}^\mathrm{dyn}$ remains roughly
constant while, for $d=3\,mm$, the variation is intermediate.
 These observations may indicate that, for $H=60\%$,
small beads are initially dragged upwards by the motion of the
tube and that rearrangements of the packing occur later. For
larger, heavier beads, this effect is smaller.

Additional clues are provided by video recordings and
spatio-temporal diagrams of the bead motion close to the wall. At
low relative humidities, they are almost motionless while, at
large ones, they move permanently. The amplitude of the motion is
larger for smaller beads and at high velocities and, in the
transition regimes, in the upper part of the tube.

\begin{figure}[t!]
\includegraphics[width=8cm]{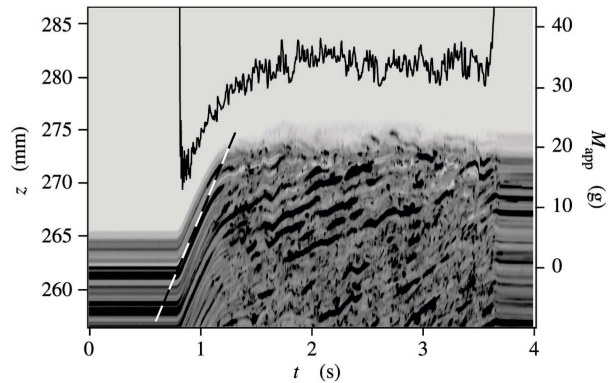} \caption{Spatio-temporal
diagram of the top of the grain packing ($z$ = distance from
bottom of the packing, slope of dashed line corresponds to a
$25\,mm\,s^{-1}$ velocity). Upper curve: simultaneous recording of
$M_\mathrm{app}$ as a function of time ($V=40\,mm\,s^{-1}$,
$d=2\,mm$, $H=60\%$).} \label{spatio}
\end{figure}

Spatio-temporal diagrams of light intensity variations on the
outside of the packing (Fig.~\ref{spatio}) are realized with a
linear CCD camera (sampling rate: 2000 lines/s). The $x$-axis
corresponds to time and the $z$-axis to distance along a vertical
line in the packing. Figure~\ref{spatio} corresponds to an
experiment during which significant bead motions occurred (note
that the acceleration of the tube is instantaneous at the scale of
the diagram). Horizontal stripes at the left and right ends of the
diagram are obtained before and after the motion. At the beginning
of the displacement, the upper surface of the packing rises
sharply (strongly inclined stripes in the diagram) and the mean
particle fraction decreases by $\Delta c \simeq 1.5\%$. After the
height of the packing has become constant, the grains still drift
upwards, but slower and a compensating downwards flow takes place
in the central part of the packing. Finally, after the tube has
stopped, the packing surface does not move down to its original
location. No clear-cut dependence of these particle fraction
variations on the tube velocity was observed.

The upper curve in Fig.~\ref{spatio} displays the variation of
$M_\mathrm{app}$ with time during that same experiment: the
variations of $M_\mathrm{app}$ lasts a little longer than those of
the mean particle fraction. This implies that rearrangements of
the force distribution still occur after the decompaction has
stopped. At low relative humidities ($H=25\%$), on the contrary,
no upwards drifts of the beads or of the packing surface are
observed.

The results presented above demonstrate that dynamical variations
of the apparent mass $M_\mathrm{app}$ are very sensitive to the
relative humidity $H$ although Janssen's model still remains
valid. This dependence is likely to correspond to variations of
the capillary forces between grains and between the grains and the
walls. Many recent studies have dealt with the dependence of these
forces on the content and distribution of liquids in granular
materials \cite{Hornbaker97, Mason99, Tegzes99, Halsey98}. Even
for small amounts of liquid at the contact point between beads,
large capillary forces appear on a single asperity or on several
asperities without requiring that pendular liquid bridges build up
over the full contact region.

At large relative humidities (typ. $H=60\%$, $80\%$), interactions
between grains and with the walls are strong enough so that grains
get dragged upwards by the tube walls (reducing strongly
$M_\mathrm{app}$ in the process) and keep moving even after a
constant value of $M_\mathrm{app}$ has been reached. A significant
decompaction of the packing occurs (by 2.5\% in our experiments)
and does not disappear after the tube has stopped. Very similar results
are obtained for $H=60\%$ and $80\%$: this implies that interaction forces
do not depend very much on $H$ once liquid bridges have built up on the rugosities.

For low relative humidities ($H=25\%$), the capillary interactions
should be weaker and take place on a few asperities: this would
explain why beads remain almost motionless during the experiment
although the interaction with the walls is strong enough to modify
the force distribution inside the packing at the onset of the
motion. Also, for single beads, the ratio of the capillary forces
and the weight of the grains is likely to be larger for smaller
beads, explaining the distinct variation for $d = 1.5\,mm$ in
Fig.~\ref{lambdaV}b. The smaller amplitude of the displacements of
the beads allows in addition to reach faster a dynamical
equilibrium configuration: this is shown by the time dependence of
$M_\mathrm{app}$ which becomes immediately constant in contrast
with higher values of $H$. Also, for $H=25\%$, the perturbation
induced by the motion increases with the final velocity $V$ of the
tube: this is shown by the linear decrease of
$M_\mathrm{app}^\mathrm{dyn}$ with $V$ and contrasts with its
constant value for $H=60\%$ and $80\%$.

Note that no electrostatic effects were observed, even for $H=25\%$.

To conclude, these experiments have allowed for a sensitive study
of the dependence of force screening effects on the relative
humidity. Increasing the latter enhances interactions between
grains and with the walls, resulting in a significant decompaction
and in a much larger amplitude of the grain motions. Janssen's
model however still describes correctly in all cases variations of
the effective mass with the height of the packing: the main
difference is a decrease of the screening length with an
increasing tube velocity for the lowest relative humidity while it
is constant at larger ones. In future works, this information will
need to be complemented by the influence of two additional
important parameters: the roughness of the tube walls and of the
beads and the mean particle fraction of the packing.

We thank Ch. Fr\'enois for the realization and development of the
motion control system, G. Chauvin and R. Pidoux for the
realization of the experimental setup, \'E. Cl\'ement and B.
Perrin for helpful discussions and Ph. Gondret for a careful
reading of the manuscript.


\end{document}